\documentclass[12pt]{article}%
\usepackage{amsmath}
\usepackage{amsfonts}
\usepackage{amssymb}
\usepackage{graphicx}%
\newcommand{\bx}{\mathbf{x}}
\newcommand{\bu}{\mathbf{u}}

\newcommand{\leftcontract}{\mathbin\lrcorner}
\newcommand{\rightcontract}{\mathbin\llcorner}

\def\b<#1>{\langle #1 \rangle}

\begin{document}

\title{The Hyperbolic Clifford Algebra of Multivecfors}
\author{{\small Waldyr A. Rodrigues Jr.} {\small and Quintino A. G. de Souza}\\$\hspace{-0.1cm}${\footnotesize Institute of Mathematics, Statistics and
Scientific Computation}\\{\footnotesize IMECC-UNICAMP CP 6065}\\{\footnotesize 13083-859 Campinas, SP, Brazil}\\{\footnotesize e-mails: walrod@ime.unicamp.br; quin@ime.unicamp.br }}
\maketitle

\begin{abstract}
In this paper we give a thoughtful exposition of the Clifford algebra
$\mathcal{C\ell(}H_{V})$ of \textit{multivecfors} which is naturally
associated with a hyperbolic space $H_{V}$, whose elements are called
\textit{vecfors}. Geometrical interpretation of vecfors and multivecfors are
given. Poincar\'{e} automorphism (Hodge dual operator) is introduced and
several useful formulas derived. The role of a particular ideal in
$\mathcal{C\ell(}H_{V})$ whose elements are representatives of spinors and
resume the \textit{algebraic} \textit{properties} of Witten superfields is discussed.

\end{abstract}

\newpage

\section{Introduction}

In this paper we give a thoughtful presentation of the hyperbolic Clifford
algebra \textit{ }of a real $n$-dimensional space $V$. \ We start recalling in
Section 2.1 the construction of a hyperbolic space $H_{V}=(V\oplus V^{\ast
},\b<\ ,\ >)$ endowed with a "duality" non-degenerate bilinear form
$\b<\ ,\ >$ and then show how to relate $H_{V}$ with the hyperbolic space
associated with the exterior direct sums of pairs $(V,b)$ and $(V,-b)$ where
$b$ \ is an arbitrary symmetric bilinear form on $V$ of arbitrary signature.
In Section 2.2. we introduce an orientation for $H_{V}$ and in Section 2.3 we
give a geometrical interpretation for the elements of $H_{V}$ that we call
\textit{vecfors}. In Section 3 we study the endomorphisms of $H_{V}$ and
introduce the concept of isotropic extensions. In Section 4.1 we introduce $(%
{\displaystyle\bigwedge}
H_{V},\b<\ ,\ >)$ the exterior algebra of \textit{multivecfors} endowed with a
canonical bilinear form $\b<\ ,\ >$ induced by $\b<\ ,\ >$. Left and right
contractions of multivecfors are introduced and their properties are detailed.
In Section 4.2 we introduce the Poincar\'{e} automorphism (Hodge dual) on $(%
{\displaystyle\bigwedge}
H_{V},\b<\ ,\ >)$ and study its properties. Section 4.3 discusses a
differential algebra structure on $%
{\displaystyle\bigwedge}
H_{V}$. In Section 5 we introduce $\mathcal{C\ell(}H_{V})$ the Clifford
algebra of multivecfors (or \textit{mother} algebra), and recall that it is
the algebra of endomorphisms of $%
{\displaystyle\bigwedge}
H_{V}$, i.e., $\mathcal{C\ell(}H_{V})\simeq\mathrm{End}(%
{\displaystyle\bigwedge}
H_{V})$. A \textit{grandmother}algebra $\mathcal{C\ell(}H_{V}^{2}%
)\simeq\mathrm{End}(\mathcal{C\ell(}H_{V}))$ where $H_{V}^{2}$ is a second
order hyperbolic structure is also discussed. In Section 6 we study the
structure of a particular minimal ideal in $\mathcal{C\ell(}H_{V})$ whose
elements (representatives of spinors) represent the \textit{algebraic
properties} of superfields. In Section 7 we present our conclusions.

\section{Hyperbolic Spaces}

\subsection{Definition; Basic Properties}

Let $V$ , $V^{\ast}$ be a pair of dual $n$-dimensional vector spaces over the
real field $\mathbb{R}$. We call hyperbolic structure\/ over $V$ to the
pair\footnote{$V\oplus V^{\ast}$ means the exterior direct sum of the vector
spaces $V$ and $V^{\ast}$.}
\[
H_{V}=(V\oplus V^{\ast},\b<\ ,\ >)
\]
where $\b<\ ,\ >$ is the non-degenerate symmetric bilinear form of index~$n$
defined by
\[
\b<\mathbf{x},\mathbf{y}>=x^{\ast}(y_{\ast})+y^{\ast}(x_{\ast})
\]
for all $\mathbf{x}=x_{\ast}\oplus x^{\ast}$, $\mathbf{y}=y_{\ast}\oplus
y^{\ast}\in H_{V}$.

The elements of $H_{V}$ will be referred as \textit{vecfors}. A vecfor
$\mathbf{x}=x_{\ast}\oplus x^{\ast}\in H_{V}$ will be said to be

(i) positive, if $x^{\ast}(x_{\ast})>0,$

(ii) null, if $x^{\ast}(x_{\ast})=0,$

(iii) negative, if $x^{\ast}(x_{\ast})<0.$

If $x^{\ast}(x_{\ast})=\pm1$, then we say that $\mathbf{x}=x_{\ast}\oplus
x^{\ast}$ is a unit vecfor.

Let $H_{V}^{\ast}$ denote the dual space of $H_{V}$, i.e., $H_{V}^{\ast}=((V$
$\oplus V^{\ast})^{\ast},\b<\ ,\ >^{-1})$, where $\b<\ ,\ >^{-1}$ is the
\textit{reciprocal} of the bilinear form $\b<\ ,\ >$. We have the following
natural isomorphisms:
\[
H_{V}^{\ast}\simeq H_{V^{\ast}}\simeq H_{V}%
\]
where $H_{V^{\ast}}$ is the hyperbolic space over $V^{\ast}$. Therefore,
$H_{V}$ is an \textquotedblleft auto-dual\textquotedblright\ space, i.e., it
may be canonically identified with its dual space.

The spaces $V$ and $V^{\ast}$ are naturally identified with their images in
$H_{V}$ under the inclusions $i_{\ast}:V$ $\rightarrow V$ $\oplus V^{\ast}$,
$x_{\ast}\mapsto i_{\ast}x_{\ast}=x_{\ast}\oplus0\equiv x_{\ast}$, and
$i^{\ast}:V^{\ast}\rightarrow V$ $\oplus V^{\ast}$, $x^{\ast}\mapsto i^{\ast
}x^{\ast}=0\oplus x^{\ast}\equiv x^{\ast}$. Then,
\[
\b<x_{\ast},y_{\ast}>=0, \quad\b<x^{\ast},y^{\ast}>=0, \quad\text{and}%
\quad\b<x^{\ast},y_{\ast}>=x^{\ast}(y_{\ast}),
\]
for all $x_{\ast},y_{\ast}\in V$ $\subset V$ $\oplus V^{\ast}$ and $x^{\ast
},y^{\ast}\in V^{\ast}\subset V$ $\oplus V^{\ast}$. This means that $V$ and
$V^{\ast}$ are maximal totally \textit{isotropic} subspaces of $H_{V}$ and any
pair of dual basis $\{e_{1},\ldots,e_{n}\}$ in $V$ and $\{\theta^{1}%
,\ldots,\theta^{n}\}$ in $V^{\ast}$ determines a \textit{Witt basis\/} in
$H_{V}$, i.e., a basis satisfying
\[
\b<e_{i},e_{j}>=0, \quad\b<\theta^{i},\theta^{j}>=0, \quad\text{and}%
\quad\b<\theta^{i},e_{j}>=\delta_{j}^{i}.
\]
\medskip

\noindent\textbf{Remark. }More generally, to each subspace $S\subset V$ (or,
analogously, $S^{\prime}\subset V^{\ast}$) we can associate a maximal totally
\textit{isotropic subspace} $I(S)\subset H_{V}$ as follows: Define the null\/
subspace $S^{\prime}\subset V^{\ast}$ of $S\subset V$ by
\[
S^{\prime}=\{x^{\ast}\in V^{\ast},\ x^{\ast}(y_{\ast})=0\ \forall y_{\ast}\in
S\}.
\]
We mention without proof that the null subspaces satisfy the properties:

(i) $S^{\prime\prime}=S,$

(ii) $S_{1}\subset S_{2}\implies S_{2}^{\prime}\subset S_{1}^{\prime},$

(iii) $(S_{1}+S_{2})^{\prime}=S_{1}^{\prime}\cap S_{2}^{\prime},$

(iv) $(S_{1}\cap S_{2})^{\prime}=S_{1}^{\prime}+S_{2}^{\prime},$

(v) $\dim S+\dim S^{\prime}=\dim V,$

(vi) $S^{\ast}=\frac{V^{\ast}}{S^{\prime}}.$

Now construct the space%
\[
I(S)=S\oplus S^{\prime}%
\]
which is a $n$-dimensional vector subspace of $H_{V}$, according to property
(v). From the definition of $S^{\prime}$, we get $\b<\mathbf{x},\mathbf{y}>=0$
for all $\mathbf{x},\mathbf{y}\in I(S)$, proving thus that $I(S)$ is a maximal
totally isotropic subspace of $H_{V}$. The null space of $I(S)$ is the space
$I^{\prime}(S)=S^{\prime}\oplus S\simeq I(S)$, whereas the dual space
$I^{\ast}(S)$ of $I(S)$ is given by the quotient%
\[
I^{\ast}(S)=\frac{H_{V}^{\ast}}{I^{\prime}(S)}\simeq\frac{H_{V}}{I(S)}%
\]
Of course, $I^{\ast}(S)$ is itself a maximal totally isotropic subspace of
$H_{V}$. The above relation also means that
\[
H_{V}\simeq I(S)\oplus I^{\ast}(S)
\]
for any $S\subset V$.

To each Witt basis $\{e_{1},\ldots,e_{n},\theta^{1},\ldots,\theta^{n}\}$ of
$H_{V}$, we can associate an orthonormal basis $\{%
\hbox{\boldmath$\sigma$}%
_{1},\ldots,%
\hbox{\boldmath$\sigma$}%
_{2n}\}$ of $H_{V}$ by letting
\[%
\hbox{\boldmath$\sigma$}%
_{k}={{\frac{1}{\sqrt{2}}}}(e_{k}\oplus\theta^{k})\qquad\text{and}\qquad%
\hbox{\boldmath$\sigma$}%
_{n+k}={{\frac{1}{\sqrt{2}}}}(\bar{e}_{k}\oplus\theta^{k})
\]
where $\bar{e}_{k}=-e_{k}$, $k=1,\ldots,n$. We have,
\[
\b<\hbox{\boldmath$\sigma$}_{k}, \hbox{\boldmath$\sigma$}_{l}>=\delta_{kl},
\quad\b< \hbox{\boldmath$\sigma$}_{n+k}, \hbox{\boldmath$\sigma$}_{n+l}
>=-\delta_{kl}, \quad\text{and}\quad\b< \hbox{\boldmath$\sigma$}_{k},
\hbox{\boldmath$\sigma$}_{n+l} >=0,
\]
$k,l=1,\ldots,n$. If $\mathbf{x}=x_{\ast}\oplus x^{\ast}$, with $x_{\ast
}=x_{\ast}^{k}e_{k}$ and $x^{\ast}=x_{k}^{\ast}\theta^{k}$, then the
components $(x^{k},x^{n+k})$ of $\mathbf{x}$ with respect to $%
\hbox{\boldmath$\sigma$}%
_{k}$ are
\begin{equation}
x^{k}={{\frac{1}{\sqrt{2}}}}(x_{k}^{\ast}+x_{\ast}^{k})\qquad and\qquad
x^{n+k}={{\frac{1}{\sqrt{2}}}}(x_{k}^{\ast}-x_{\ast}^{k}). \label{Eq-1}%
\end{equation}

The base vectors $%
\hbox{\boldmath$\sigma$}%
_{n+k}$ are obtained from the $%
\hbox{\boldmath$\sigma$}%
_{k}$ through the involution $\mathbf{x}=x_{\ast}\oplus x^{\ast}\mapsto
\bar{\mathbf{x}}$, where
\[
\bar{\mathbf{x}}=(-x_{\ast})\oplus x^{\ast}%
\]
is the (hyperbolic) \textit{conjugate}\/ of the vecfor $\mathbf{x}$. Note that
$\bar{\mathbf{x}}$ is orthogonal to $\mathbf{x}$ and that its square has
opposite sign to that of $\mathbf{x}$, e.g.,
\[
\b<\bar{\mathbf{x}},\mathbf{x}>=0 \qquad\text{and}\qquad\b<\bar{\mathbf{x}%
},\bar{\mathbf{x}}>=-\b<\mathbf{x},\mathbf{x}>.
\]
With the components of $\mathbf{x}$ given by Eq.~\ref{Eq-1}, the components
$(\bar{x}^{k},\bar{x}^{n+k})$ of $\bar{\mathbf{x}}$ in the basis $\{%
\hbox{\boldmath$\sigma$}%
_{k}\}$ would be given by
\[
\bar{x}^{k}={{\frac{1}{\sqrt{2}}}}(x_{k}^{\ast}-x_{\ast}^{k})\qquad
\text{and}\qquad\bar{x}^{n+k}={{\frac{1}{\sqrt{2}}}}(x_{k}^{\ast}+x_{\ast}%
^{k}).
\]

\noindent\textbf{Remark. }\bigskip The hyperbolic conjugation $\overline
{}:H_{V}\rightarrow H_{V}$ together with the canonical bilinear form
$\langle\;,\;\rangle$, enable us to define a bilinear form $[\;,\;]$ by
\[
\lbrack\mathbf{x,y]=}\langle\;\mathbf{\bar{x}},\mathbf{y}\;\rangle
\]
for all $\mathbf{x,y}\in H_{V}$. For $\mathbf{x}=x_{\ast}+x^{\ast}$ and
$\mathbf{y}=y_{\ast}+y^{\ast}$, it holds
\[
\lbrack\mathbf{x,y]=}x^{\cdot}(y_{\ast})+y^{\ast}(x_{\ast})
\]
and it follows that
\[
\lbrack\mathbf{x,y]}=-[\mathbf{y,x]}
\]
meaning that $[\;,\;]$ is an antisymmetric bilinear form on $H_{V}$.

Due to the auto-duality of $H_{V}$ stated by the isomorphisms $H_{V}^{\ast
}\simeq H_{V^{\ast}}\simeq H_{V}$, we can use the notation $\{%
\hbox{\boldmath$\sigma$}%
^{1},\ldots%
\hbox{\boldmath$\sigma$}%
^{2n}\}$ to indicate the dual basis $\{%
\hbox{\boldmath$\sigma$}%
_{1},\ldots%
\hbox{\boldmath$\sigma$}%
_{2n}\}$ as well the reciprocal basis of this same basis. As elements of
$H_{V}^{\ast}$ the expressions of the $%
\hbox{\boldmath$\sigma$}%
^{k\prime}$s are%
\[%
\hbox{\boldmath$\sigma$}%
^{k}=\frac{1}{\sqrt{2}}\left(  \theta^{k}+e_{k}\right)  \text{ and\ }%
\hbox{\boldmath$\sigma$}%
_{k}=\frac{1}{\sqrt{2}}\left(  \bar{\theta}^{k}+e_{k}\right)  .
\]

Another remarkable result on the theory of hyperbolic spaces is the following

\noindent\textbf{Proposition 1. }Given an arbitrary non-degenerate symmetric
bilinear form $b$ on $V$, there is an isomorphism
\[
H_{V}\simeq(V,b)\oplus(V,-b)=H_{bV}
\]

\noindent\textbf{Proof. }See,e.g., \cite{Knus}. The isomorphism is given by
the mapping $\rho_{b}:H_{V}\rightarrow H_{bV}$, by $x_{\ast}+x^{\ast}\mapsto
x_{+}+x_{-}$, where
\[
x_{\pm}=\frac{1}{\sqrt{2}}\left(  b^{\ast}x^{\ast}\pm x_{\ast}\right)
\]
with $b^{\ast}x^{\ast}:=b^{\ast}(x^{\ast},\;)$ for all $x^{\ast}\in V^{\ast}$,
\ \ where $b^{\ast}$, the \textit{reciprocal} of $b$ is a bilinear form in
$V^{\ast}$ such that $b^{ik}b_{kj}=\delta_{j}^{i}$, $b_{ij}=b(e_{i},e_{j})$,
$b^{ik}=b^{\ast}(\theta^{i},\theta^{k})$.

Observe that the image of the basis $\{%
\hbox{\boldmath$\sigma$}%
_{1},\ldots%
\hbox{\boldmath$\sigma$}%
_{2n}\}$ of $H_{V}$ under the above isomorphism is the basis $\{\mathbf{e}%
_{1},...,\mathbf{e}_{2n}\}$ of $H_{bV}$ given by
\[
\mathbf{e}_{k}=\frac{1}{\sqrt{2}}\left[  \left(  e_{k}+e^{k}\right)
\oplus(e_{k}-e^{k})\right]  ,
\]
and
\[
\mathbf{e}_{n+k}=\frac{1}{\sqrt{2}}\left[  \left(  e_{k}-e^{k}\right)
\oplus(e_{k}+e^{k})\right]  ,
\]
where $e^{k}=b^{kl}e_{l}$, $k=1,2,\ldots,n$. The basis $\{e_{k}\}$ need not to
be a $b$-orthonormal basis of $V$; but in case it is, the above expressions
become:
\[
\mathbf{e}_{k}=\left\{
\begin{array}
[c]{ccc}%
e_{k}\oplus0, &  & k=1,\ldots,p\\
0\oplus e_{k} &  & k=p+1,\ldots,n
\end{array}
\right.  ,
\]
and
\[
\mathbf{e}_{n+k}=\left\{
\begin{array}
[c]{ccc}%
0\oplus e_{k}, &  & k=1,\ldots,p\\
e_{k}\oplus0 &  & k=p+1,\ldots,n
\end{array}
\right.  ,
\]
where $p$ is the index of $b$.

\textbf{\noindent Remark}. The hyperbolic structure $H_{V\oplus V^{\prime
}\text{ }}$of the direct sum of two vector spaces $V$ and $V^{^{\prime}}$
satisfies
\[
H_{V\oplus V^{\prime}\text{ }}\simeq H_{V\text{ }}\oplus H_{V^{\prime}\text{
}},
\]
where $\langle\;,\;\rangle_{^{V\oplus V\prime}}=\langle\;,\;\rangle_{V}%
\oplus\langle\;,\;\rangle_{V^{\prime}}$. Taking $V^{\prime}=V^{\ast}$and
writing $H_{V}^{2}=H_{H_{V}}$ we conclude that
\[
H_{V}^{2}\simeq H_{V\text{ }}\oplus H_{V^{\prime}\text{ }}.
\]

Moreover, it follows from Proposition 1, reminding that $H_{V}$ is a metric
space, that
\[
H_{V}^{2}\simeq H_{V\text{ }}\oplus H_{-V\text{ }},
\]
where $H_{-V\text{ }}=\left(  V\oplus V^{\ast},-\langle\;,\;\rangle\right)  $.
$H_{V}^{2}$ is called \textit{second order} hyperbolic space.

To a pair of dual basis $\{e_{1},\ldots,e_{n}\}$ in $V$ and $\{\theta
^{1},\ldots,\theta^{n}\}$ in $V^{\ast}$, we associate an orthonormal basis
$\{\Sigma_{1},\ldots,\Sigma_{4n}\}$ of $H_{V}^{2}$ by the relations
\begin{align*}
\Sigma_{k}  &  ={\frac{1}{\sqrt{2}}}%
\hbox{\boldmath$\sigma$}%
_{k}\oplus%
\hbox{\boldmath$\sigma$}%
^{k}\\
\Sigma_{n+k}  &  ={\frac{1}{\sqrt{2}}}%
\hbox{\boldmath$\sigma$}%
_{n+k}\oplus%
\hbox{\boldmath$\sigma$}%
^{n+k}\\
\Sigma_{2n+k}  &  ={\frac{1}{\sqrt{2}}}(-%
\hbox{\boldmath$\sigma$}%
_{k})\oplus%
\hbox{\boldmath$\sigma$}%
^{k}\\
\Sigma_{3n+k}  &  ={\frac{1}{\sqrt{2}}}(-%
\hbox{\boldmath$\sigma$}%
_{n+k})\oplus%
\hbox{\boldmath$\sigma$}%
^{n+k}%
\end{align*}
$k=1,\ldots,n$, where $\{%
\hbox{\boldmath$\sigma$}%
_{k}\}$ and $\{%
\hbox{\boldmath$\sigma$}%
^{k}\}$ are like before.

\subsection{\textrm{Orientation}}

Besides a canonical metric structure, a hyperbolic space is also provided with
a canonical sense of orientation, induced by the $2n$-vector $%
\hbox{\boldmath$\sigma$}%
\in\bigwedge\nolimits^{2n}H_{V}$ given by
\[%
\hbox{\boldmath$\sigma$}%
=%
\hbox{\boldmath$\sigma$}%
_{1}\wedge\ldots\wedge%
\hbox{\boldmath$\sigma$}%
_{2n}%
\]
where $\{%
\hbox{\boldmath$\sigma$}%
_{1},\ldots,%
\hbox{\boldmath$\sigma$}%
_{2n}\}$ is the orthonormal basis of $H_{V}$ naturally associated with the
dual basis $\{e_{1},\ldots,e_{n}\}$ of $V$ and $\{\theta^{1},\ldots,\theta
^{n}\}$ of $V^{\ast}$. Note that $%
\hbox{\boldmath$\sigma$}%
$ satisfies
\[
\b<\hbox{\boldmath$\sigma$},\hbox{\boldmath$\sigma$}>=(-1)^{n}.
\]
Here, $\b<\ ,\ >$ denotes the natural extension \cite{rodoliv2006} of the
metric of $H_{V}$ to the space $\bigwedge^{2n}H_{V}$.

The reason for saying that $%
\hbox{\boldmath$\sigma$}%
$ defines a canonical sense of orientation is that it is independent on the
choice of the basis for $V$. To see this it is enough to verify that
\[
\hbox{\boldmath$\sigma$}%
=e_{\ast}\wedge\theta^{\ast}
\]
where\textrm{\ }%
\[
e_{\ast}=e_{1}\wedge\ldots\wedge e_{n}\qquad and\qquad\theta^{\ast}=\theta
^{1}\wedge\ldots\wedge\theta^{n}
\]
Thus, under a change of basis in $V$ , $e_{\ast}$ transforms as $\lambda
e_{\ast}$, $\lambda\neq0$, whereas $\theta^{\ast}$ transforms as $\lambda
^{-1}\theta^{\ast}$, so that $%
\hbox{\boldmath$\sigma$}%
$ remains unchanged.

\subsection{Geometrical Interpretation of Vecfors}

We start recalling that, with respect to its underlying vector space, every
vector is represented by an arrow (oriented line segment) pointing outward the
origin (zero vector), with the operations of multiplication by a scalar and
the operation of sum of vectors by the parallelogram rule having their usual interpretation.

However, it is often useful or necessary to interpret the elements of a given
vector space with respect to another one,\textrm{\ }related to the first by
some prescription. In particular, our aim here is to show how the elements of
the hyperbolic space $H_{V}$ are represented taking the space $V$ as the
\textquotedblleft basis\textquotedblright\ for the representation. That is, we
attribute to elements of $V$ their usual interpretation as arrows and we want
to describe elements of $H_{V}$ as collections of such arrows.

Since vecfors are pairs constituted by a vector and a linear form, the first
question is how a linear form $\alpha\in V^{\ast}$ is represented within $V$ .
The answer is that it is described as an ordered pair of $(n-1)$-dimensional
parallel hyperplanes $(S_{0}(\alpha),S_{1}(\alpha))$, given by
\[
S_{0}(\alpha)=\{x\in V,\ \alpha(x)=0\} \quad\text{and} \quad S_{1}%
(\alpha)=\{x\in V,\ \alpha(x)=1\}
\]
It should be noted that only $S_{0}(\alpha)$, which indeed is the null
subspace spanned by $\alpha$, is a vector subspace of $V$ . The notation
$S_{a}(\alpha)$, $a\in\mathbb{R}$, will be used here to indicate the set of
vectors $x\in V$ satisfying the equation $\alpha(x)=a$. The sequence
$\{S_{a}(\alpha),\ a\in\mathbb{R}\}$ can also be used as an alternative
representation of a linear form.

If $\alpha^{\prime}=a\alpha$, $a\neq0$, we have $S_{0}(\alpha^{\prime}%
)=S_{0}(\alpha)$, while
\begin{align*}
S_{1}(\alpha^{\prime})  &  =\{x\in V,\ \alpha^{\prime}(x)=1\}\\
&  =\{x\in V,\ \alpha(x)=1/a\}\\
&  =\{x\in V,\ x=y/a,\ y\in S_{1}(\alpha)\}
\end{align*}
which denotes the affine $(n-1)$-dimensional hyperplane obtained from
$S_{1}(\alpha)$ by \textquotedblleft dilating\textquotedblright\ each vector
of $S_{1}(\alpha)$ by a factor $1/a$. If $a>1$, such a dilation is indeed a
contraction: multiplying a linear form by a factor greater than one yields the
linear form represented by affine hyperplanes parallel to those representing
the original linear form, but which are closer from each other, with respect
to the original hyperplanes, by the factor $1/a$. Analogously, for\textrm{\ }%
$0<a<1$, the hyperplanes representing $\alpha^{\prime}$ will be separated from
each other, with respect to the hyperplanes representing $\alpha$, by a factor
$1/a>1$, thus representing a \textquotedblleft true\textquotedblright%
\ dilation in $V$\textrm{\ . }

For\textrm{\ }$a=-1$, the hyperplane\textrm{\ }$S_{1}(\alpha^{\prime}%
)$\textrm{\ }will be constituted by the negatives of the vectors in
$S_{1}(\alpha)$,\textrm{\ }so that both hyperplanes will be separated from
$S_{0}(\alpha)$\textrm{\ }by the same amount, but will be given by
the\textrm{\ }(\textquotedblleft anti-\textquotedblright) mirror image of each other.

Given now\textrm{\ }$\alpha,\beta\in V^{\ast}$\textrm{, }their sum\textrm{\ }%
$\alpha+\beta\in V^{\ast}$\textrm{\ }is represented in\textrm{\ }%
$V$\textrm{\ }as follows: The hyperplane $S_{0}(\alpha+\beta)$\textrm{\ }is
given by the unique\textrm{\ }$(n-1)$-dimensional affine hyperplane containing
the intersections\textrm{\ }$S_{1}(\alpha)\cap S_{1}(-\beta)$ and
$S_{1}(-\alpha)\cap S_{1}(\beta)$\textrm{\ }(note that these
intersections\textrm{\ }are $(n-2)$-dimensional affine hyperplanes,
unless\textrm{\ }$\alpha=\beta$\textrm{\ }when we\textrm{\ }have\textrm{\ }%
$S_{0}(\alpha+\beta)=S_{0}(\alpha)$). In turn,\textrm{\ }$S_{1}(\alpha+\beta
)$\textrm{\ }is given by the unique hyperplane containing the
intersections\textrm{\ }$S_{0}(\alpha)\cap S_{1}(\beta)$\textrm{\ }%
and\textrm{\ }$S_{0}(\beta)\cap S_{1}(\alpha)$. Observe also that we can take
advantage from the\textrm{\ }fact that the hyperplanes representing a linear
form are parallel and to obtain\textrm{\ }$S_{0}(\alpha+\beta)$ as the unique
hyperplane containing the origin which is parallel to $S_{1}(\alpha+\beta)$.

With the above constructions, the geometrical meaning of the vecfors become
apparent: an element\textrm{\ }$\mathbf{x}=x\oplus\alpha\in H_{V}$%
\textrm{\ }is represented by an arrow \textrm{\ }$\vec{x}$\textrm{\ \ }%
together with a pair of hyperplanes\textrm{\ }$S_{0}(\alpha)$ and
$S_{1}(\alpha)$. The sum of two vecfors is obtained geometrically by summing
up independently the representations of their vector parts (according to the
parallelogram rule) and the representations of their linear form parts
(according to the sum rule stated above)\textrm{. }The multiplication of a
vecfor $\mathbf{x}$ by a scalar $a$, in turn, has the interesting property
that its vector part is represented by an arrow \textquotedblleft
dilated\textquotedblright\ by a factor $a$ with respect to the original one,
while the hyperplanes representing its linear form part are \textquotedblleft
contracted\textquotedblright\ by the same factor with respect to the original one.

We had classified a vecfor $\mathbf{x}=x\oplus\alpha$ by the value of the
contraction $\alpha(x)$. Null vecfors were defined by the condition
$\alpha(x)=0$, which is the defining condition of the hyperplane $S_{0}%
(\alpha)$. Therefore, null vecfors stand geometrically for vecfors such that
the vector part is \textquotedblleft parallel\textquotedblright\ to the linear
form part.

For a non-null vecfor, the contraction $\alpha(x)$ gives the amount by which
the vector $x$ should be contracted, and possibly reflected, in order to get a
\textquotedblleft unitary\textquotedblright\ vector with respect to $\alpha$,
i.e., a vector with its tip lying on the hyperplane $S_{1}(\alpha)$. Positive
vecfors, for which $\alpha(x)>0$, are such that their vector part have the
same \textquotedblleft orientation\textquotedblright\ of their linear form
part, so that only a \textquotedblleft contraction\textquotedblright\ need to
be done in order to normalize them. In turn, the vector parts of negative
vecfors, have opposite \textquotedblleft orientation\textquotedblright\ of
those of their linear form parts, so that they need to be reflected through
the origin, in addition to be contracted, in order to be normalized.

\begin{figure}[tb]
\unitlength 1mm
\par
\begin{center}
\textrm{\begin{picture}(30,25)
\put(0,10){\line(1,0){30}}
\put(0,20){\line(1,0){30}}
\put(15,10){\vector(0,1){15}}
\put(30,22){\vector(0,1){0}}
\put(15,9){\makebox(0,0)[t]{$0$}}
\end{picture}
\hfil   \begin{picture}(30,25)
\put(0,10){\line(1,0){30}}
\put(0,20){\line(1,0){30}}
\put(22.5,15){\vector(-1,0){15}}
\put(30,22){\vector(0,1){0}}
\put(15,10){\makebox(0,0){$\cdot$}}
\put(15,9){\makebox(0,0)[t]{$0$}}
\end{picture}
\hfil   \begin{picture}(30,25)
\put(0,10){\line(1,0){30}}
\put(0,20){\line(1,0){30}}
\put(15,10){\vector(1,-1){10}}
\put(30,22){\vector(0,1){0}}
\put(15,11){\makebox(0,0)[b]{$0$}}
\end{picture}
}
\end{center}
\caption{Geometrical representations of (a)~a positive, (b)~a null, and (c)~a
negative vecfor}%
\end{figure}
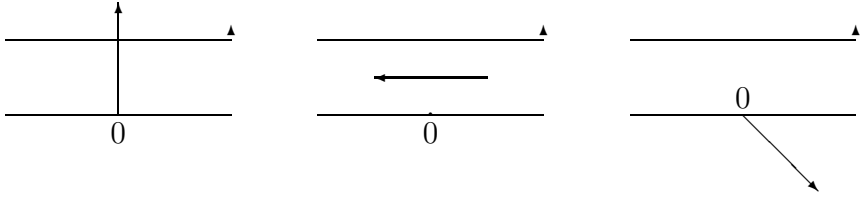

\section{Endomorphisms of $H_{V}$}

\subsection{Isotropic Extensions}

Given a linear mapping $\phi:V$ $\rightarrow V$ (i.e., $\phi\in
\mathop{\mathrm{End}} V$ ), we define $\phi^{\ast}:V^{\ast}\leftarrow V^{\ast
}$, the dual mapping\/ of $\phi$, by
\[
(\phi^{\ast}\alpha)x=\alpha(\phi x)
\]
for all $x\in V$ and $\alpha\in V^{\ast}$. The operation $\phi\mapsto
\phi^{\ast}$, $\phi\in\mathop{\mathrm{End}} V$ is linear and satisfy

(i) $\phi^{\ast\ast}=\phi,$

(ii) $(\phi\psi)^{\ast}=\psi^{\ast}\phi^{\ast},$

(iii) $\ker\phi^{\ast}=(\mathrm{im}$ $\phi)^{\prime}$, $\mathrm{im}$
$\phi^{\ast}=(\ker\phi)^{\prime},$

(iv) $\det\phi^{\ast}=\det\phi$, $\mathrm{tr}$ $\phi^{\ast}=\mathrm{tr}$
$\phi,$

(v) $\phi(S)\subset S\implies\phi^{\ast}(S^{\prime})\subset S^{\prime}$,
$S\subset V.$

Every linear mapping $\phi\in\mathop{\mathrm{End}} V$ (or, equivalently,
$\phi^{\ast}\in\mathop{\mathrm{End}} V^{\ast}$) induces a linear mapping
$I(\phi)\in\mathop{\mathrm{End}} H_{V}$, defined by
\[
I(\phi)=\phi\oplus\phi^{\ast}.
\]
We call $I(\phi)$ isotropic extension\/ of $\phi$ ($\phi^{\ast}$) to $H_{V}$.
The reason for this nomenclature is that, due to the fifth property above, if
$S\subset V$ is stable under $\phi$, then $I(S)$ is stable under $I(\phi)$,
i.e.,
\[
I(\phi)(S\oplus S^{\prime})\subset S\oplus S^{\prime}.
\]
\medskip

\noindent\noindent\textbf{Example.} Vecfors and Endomorphisms.\textrm{\ }%
\noindent To each vecfor $\mathbf{x}=x_{\ast}\oplus x^{\ast}\in H_{V}$
corresponds a linear mapping $\mathbf{x}:V$ $\rightarrow V$ , given by
\[
\mathbf{x}(y_{\ast})=x^{\ast}(y_{\ast})x_{\ast},
\]
for all $y_{\ast}\in V$ . The dual $\mathbf{x}^{\ast}$ of $\mathbf{x}$ is
given, obviously, by
\[
\mathbf{x}^{\ast}(y^{\ast})=y^{\ast}(x_{\ast})x^{\ast},
\]
for all $y^{\ast}\in V^{\ast}$.\medskip\ 

\noindent\textbf{Example}. Hyperbolic Projections. We can associate to a
non-null vecfor $\mathbf{x}=x_{\ast}\oplus x^{\ast}$ a projection operators
$P_{\mathbf{x}}:V$ $\rightarrow V$ and $P^{\mathbf{x}}:V^{\ast}\leftarrow
V^{\ast}$ by the relations
\[
P_{\mathbf{x}}y_{\ast}=\frac{x^{\ast}y_{\ast}}{x^{\ast}x_{\ast}}x_{\ast}%
\qquad\text{and}\qquad P^{\mathbf{x}}y^{\ast}=\frac{y^{\ast}x_{\ast}}{x^{\ast
}x_{\ast}}x^{\ast}.
\]
The operator $P^{\mathbf{x}}$ is the dual of the operator $P_{\mathbf{x}}$.
Indeed, $(P_{\mathbf{x}}^{\ast}y^{\ast})(y_{\ast})=y^{\ast}(P_{\mathbf{x}%
}y_{\ast})=(x^{\ast}y_{\ast}/x^{\ast}x_{\ast})y^{\ast}(x_{\ast})=(y^{\ast
}x_{\ast}/x^{\ast}x_{\ast})x^{\ast}y_{\ast}=(P^{\mathbf{x}}y^{\ast})(y_{\ast
})$ for all $y_{\ast}\in V$ and $y^{\ast}\in V^{\ast}$.

The vector $P_{\mathbf{x}}y_{\ast}$ is given geometrically by the intersection
of the line containing the vector $x_{\ast}$ with the $S_{1}$-hyperplane
associated \ with the linear form $x^{\ast}/x^{\ast}(y_{\ast})$, while
$P^{\mathbf{x}}y^{\ast}$ is the linear form parallel to $x^{\ast}$, whose
$S_{1}$-hyperplane contains the intersection of $S_{1}(y^{\ast})$ with the
vector $x_{\ast}$.

\begin{figure}[tb]
\unitlength 1mm
\par
\begin{center}
\textrm{\begin{picture}(40,20)
\put(0,0){\line(1,0){40}}
\put(0,10){\line(1,0){40}}
\put(39,12){\vector(0,1){0}}
\put(41,11){\makebox(0,0)[l]{$x^*$}}
\put(20,0){\makebox(0,0){$\cdot$}}
\put(20,0){\vector(0,1){20}}
\put(21,21){\makebox(0,0)[b]{$x_*$}}
\put(20,0){\vector(2,3){10}}
\put(30,16){\makebox(0,0)[lb]{$y_*$}}
\multiput(0,15)(3,0){13}{\line(1,0){2}}
\put(20,15){\vector(0,1){0}}
\put(19,14){\makebox(0,0)[rt]{$P_x y_*$}}
\end{picture}
\hfil   \begin{picture}(40,30)
\put(38,0){\line(-4,1){38}}
\put(39,6){\line(-4,1){38}}
\put(39,8){\vector(1,4){0}} \put(40,7){\makebox(0,0)[lb]{$y^*$}}
\put(0,5){\line(1,0){40}}
\put(0,15){\line(1,0){40}}
\put(39,17){\vector(0,1){0}}  \put(41,16){\makebox(0,0)[l]{$x^*$}}
\put(18,5){\vector(0,1){20}} \put(18,26){\makebox(0,0)[b]{$x_*$}}
\put(39,13){\vector(0,1){0}} \put(41,11,2){\makebox(0,0)[l]{$P^x y^*$}}
\thicklines
\put(0,11.2){\line(1,0){40}}
\end{picture}
}
\end{center}
\caption{Geometrical representation of the projections induced by
hypervectors: (a)~$P_{x}y_{\ast}$ is the vector obtained from the intersection
of $x_{\ast}$ with the $S_{1}$-hyperplane of the linear form $x^{\ast}%
/x^{\ast}(y_{\ast})$; (b)~$P^{x}y^{\ast}$ is the linear form parallel to
$x^{\ast}$ whose $S_{1}$-hyperplane crosses the intersection of $x_{\ast}$
with the $S_{1}$-hyperplane of $y^{\ast}$.}%
\end{figure}
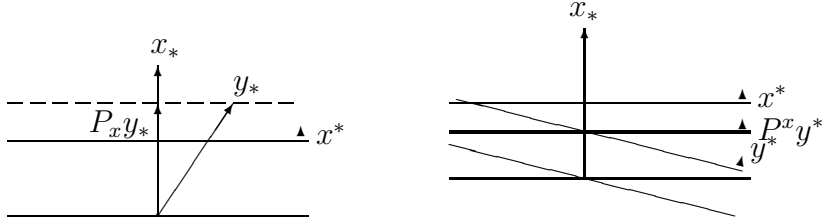

The isotropic extension of $P_{\mathbf{x}}$ to $H_{V}$, $I(P_{\mathbf{x}%
})=\mathbf{P}_{\mathbf{X}}$, is given by
\[
\mathbf{P}_{\mathbf{x}}=P_{\mathbf{x}}\oplus P^{\mathbf{x}}.
\]
The operator $\mathbf{P}_{\mathbf{x}}$ is itself a projection, since
$\mathbf{P}_{\mathbf{x}}^{2}=P_{\mathbf{x}}^{2}\oplus(P^{\mathbf{x}}%
)^{2}=P_{\mathbf{x}}\oplus P^{\mathbf{x}}=\mathbf{P}_{\mathbf{x}}$. Moreover
$\mathbf{P}_{\mathbf{x}}$ is self-dual with respect to the bilinear form
$\b<\ ,\ >$, in the sense that
\[
\b<\mathbf{P}_{\mathbf{x}}\mathbf{y},\mathbf{z}>=\b<\mathbf{y},\mathbf{P}%
_{\mathbf{x}}\mathbf{z}>.
\]
In other words, if $\mathbf{P}^{\mathbf{x}}$ denotes the dual of the
transformation $\mathbf{P}_{\mathbf{x}}$, then
\[
\mathbf{P}^{\mathbf{x}}=\mathbf{P}_{\mathbf{x}}.
\]
For the orthonormal basis $\{%
\hbox{\boldmath$\sigma$}%
_{1},\ldots,%
\hbox{\boldmath$\sigma$}%
_{2n}\}$ related to a pair of dual basis $\{e_{1},\ldots,e_{n}\}$ in $V$ and
$\{\theta^{1},\ldots,\theta^{n}\}$ in $V^{\ast}$, the matrix representations
of the projection operators $\mathbf{P}_{%
\hbox{\boldmath$\sigma$}%
_{k}}$ and $\mathbf{P}_{%
\hbox{\boldmath$\sigma$}%
_{n+k}}$ are
\[
(\mathbf{P}_{%
\hbox{\boldmath$\sigma$}%
_{k}})=(\mathbf{P}_{%
\hbox{\boldmath$\sigma$}%
_{n+k}})=\mathop{\mathrm{diag}}(0,\ldots,0,1,0,\ldots,0,1,0,\ldots,0)
\]
with the factors `$1$' appearing in the positions $k$ and $n+k$ of the
diagonal.\medskip\hfill\ 

\noindent\textbf{Example}. Hyperbolic Reflections. Reflections form another
important class of transformations in the theory of linear spaces. And it
turns out that every non-null vecfor $\mathbf{x}\in H_{V}$ can also be
associated with a reflection $R_{\mathbf{x}}:V$ $\rightarrow V$ and with a
reflection $R^{\mathbf{x}}:V^{\ast}\leftarrow V^{\ast}$ given respectively by
\[
R_{\mathbf{x}}y_{\ast}=y_{\ast}-2\frac{x^{\ast}(y_{\ast})}{x^{\ast}(x_{\ast}%
)}x_{\ast}\quad\text{and}\quad R^{\mathbf{x}}y^{\ast}=y^{\ast}-2\frac{y^{\ast
}(x_{\ast})}{x^{\ast}(x_{\ast})}x^{\ast}%
\]
for all $y_{\ast}\in V$ and $y^{\ast}\in V^{\ast}$, where $\mathbf{x}=x_{\ast
}\oplus x^{\ast}\in H_{V}$. Once again, the mappings $R_{\mathbf{x}}$ and
$R^{\mathbf{x}}$ are duals of each other.

\begin{figure}[bh]
\unitlength 1mm
\par
\begin{center}
\textrm{\begin{picture}(40,35)
\put(0,15){\line(1,0){40}}
\put(0,25){\line(1,0){40}}
\put(39,27){\vector(0,1){0}}
\put(20,15){\vector(0,1){20}}
\put(20,15){\vector(1,1){15}}
\put(20,15){\vector(1,-1){15}}
\multiput(20,30)(2,0){8}{\line(1,0){1}}
\multiput(20,0)(2,0){8}{\line(1,0){1}}
\multiput(20,0)(0,2){8}{\line(0,1){1}}
\put(20,30){\vector(0,1){0}}
\put(20,0){\vector(0,-1){0}}
\multiput(35,0)(0,2){15}{\line(0,1){1}}
\put(41,25){\makebox(0,0)[l]{$x^*$}}
\put(20,36){\makebox(0,0)[b]{$x_*$}}
\put(36,31){\makebox(0,0)[lb]{$y_*$}}
\put(36,-1){\makebox(0,0)[lt]{$R_x y_*$}}
\end{picture}
\hfil   \begin{picture}(50,35)
\put(0,15){\line(1,0){50}}
\put(0,25){\line(1,0){50}}
\put(49,27){\vector(0,1){0}}
\put(20,15){\vector(0,1){20}}
\put(4.5,27){\line(4,-3){34}}
\put(10.5,35){\line(4,-3){34}}
\multiput(0,9)(2,0){25}{\line(1,0){1}}
\put(52,25){\makebox(0,0)[l]{$x^*$}}
\put(20,36){\makebox(0,0)[b]{$x_*$}}
\put(12.7,35.3){\vector(3,4){0}}
\put(9.5,35){\makebox(0,0)[r]{$y^*$}}
\put(17.2,-.3){\vector(2,-3){0}}
\put(15,1){\makebox(0,0)[r]{$R^x y^*$}}
\thicklines
\put(16,1){\line(3,2){32}}
\put(11,9){\line(3,2){32}}
\end{picture}
}
\end{center}
\end{figure}

The isotropic extension of $R_{\mathbf{x}}$ to the space $H_{V}$,
$I(R_{\mathbf{x}})=\mathbf{R}_{\mathbf{x}}$ is given by
\[
\mathbf{R}_{\mathbf{x}}=R_{\mathbf{x}}\oplus R^{\mathbf{x}}
\]
and it should be noted that $\mathbf{R}_{\mathbf{x}}$ is an orthogonal mapping
with respect to the bilinear form $\b<\ ,\ >$ of $H_{V}$, i.e.,
\[
\b<\mathbf{R}_{\mathbf{x}}\mathbf{y},\mathbf{R}_{\mathbf{x}}\mathbf{z}%
>=\b<\mathbf{y},\mathbf{z}>
\]
for all $\mathbf{y},\mathbf{z}\in H_{V}$. This relation can also be written
as
\[
\mathbf{R}^{\mathbf{x}}\mathbf{R}_{\mathbf{x}}=\mathbf{1}
\]
with $\mathbf{R}^{\mathbf{x}}$ denoting the dual of $\mathbf{R}_{\mathbf{x}}$
and $1$ being the identity mapping of $H_{V}$.

With respect to our basis $\{%
\hbox{\boldmath$\sigma$}%
_{1},\ldots,%
\hbox{\boldmath$\sigma$}%
_{2n}\}$, the matrix representations of the mappings $\mathbf{R}_{%
\hbox{\boldmath$\sigma$}%
_{k}}$ and $\mathbf{R}_{%
\hbox{\boldmath$\sigma$}%
_{n+k}}$ are
\[
(\mathbf{R}_{%
\hbox{\boldmath$\sigma$}%
_{k}})=(\mathbf{R}_{%
\hbox{\boldmath$\sigma$}%
_{n+k}})=\mathop{\mathrm{diag}}(1,\ldots,1,-1,1,\ldots,1,-1,1,\ldots,1)
\]
where the factors `$-1$' appears in positions $k$ and $n+k$ of the
diagonal.\hfill\ 

\section{Exterior Algebra of a Hyperbolic Space}

The exterior algebra $\bigwedge H_{V}$ of the hyperbolic structure $H_{V}$ is
the pair
\[
\bigwedge H_{V}=(\bigwedge(V\oplus V^{\ast}),\b<\ ,\ >)
\]
where
\[
\bigwedge(V\oplus V^{\ast})=\sum_{r=0}^{2n}\bigwedge\nolimits^{r}(V\oplus
V^{\ast})
\]
is the exterior algebra of $V$ $\oplus V^{\ast}$ and $\b<\ ,\ >$ is the
canonical bilinear form on $\bigwedge(V$ $\oplus V^{\ast})$ induced by the
bilinear form $\b<\ ,\ >$ of $H_{V}$, e.g., for simple elements $\mathbf{x}%
_{1}\wedge\ldots\wedge\mathbf{x}_{r},\mathbf{y}_{1}\wedge\ldots\wedge
\mathbf{y}_{r}\in\bigwedge\nolimits^{r}H_{V}$, $\b<\ ,\ >^{\wedge}$ is given
by
\[
\b<\mathbf{x}_{1}\wedge\ldots\wedge\mathbf{x}_{r},\mathbf{y}_{1}\wedge
\ldots\wedge\mathbf{y}_{r}>=\mathrm{Det}\left(
\begin{array}
[c]{ccc}%
\b<\mathbf{x}_{1}, \mathbf{y}_{1}> & \ldots & \b<\mathbf{x}_{1}\ ,\mathbf{y}%
_{r}>\\
. & \ldots & .\\
. & \ldots & .\\
\b<\mathbf{x}_{r}, \mathbf{y}_{1}> & \ldots & \b<\mathbf{x}_{r}\ ,\ \mathbf{y}%
_{r}>
\end{array}
\right)  ,
\]
and it is extended by linearity and orthogonality to all of the algebra
$\bigwedge H_{V}$. Of course, due to the isomorphisms $H_{V}^{\ast}$ $\simeq
H_{V^{\ast}}\simeq H_{V}$, we have
\[
\bigwedge H_{V^{\ast}}\simeq\bigwedge H_{V}^{\ast}\simeq\bigwedge H_{V},
\]
and it follows that
\[
\bigwedge H_{V}\simeq\left(  \bigwedge H_{V}\right)  ^{\ast},
\]
i.e., $\bigwedge H_{V}$ is itself an \textit{auto-dual} space. The elements of
$\bigwedge H_{V}$ will be called \textit{multivecfors}.

Grade \textit{involution}, \textit{reversion} and \textit{conjugation} in the
algebra $\bigwedge H_{V}$ are defined as usual\footnote{See, e.g.,
\cite{rodoliv2006}.}: For a homogeneous mutivecfor $\mathbf{u}\in\bigwedge
^{r}H_{V}$,
\[
\hat{\mathbf{u}}=(-1)^{r}\mathbf{u},\qquad\tilde{\mathbf{u}}=(-1)^{\frac{1}%
{2}r(r-1)}\mathbf{u},\qquad\bar{\mathbf{u}}=(-1)^{\frac{1}{2}r(r+1)}\mathbf{u}%
\]
and we recall that every element $\mathbf{u}\in\bigwedge H_{V}$ is uniquely
decomposed into a sum
\[
\mathbf{u}=\mathbf{u}_{+}+\mathbf{u}_{-}%
\]
with $\mathbf{u}_{+}$ and $\mathbf{u}_{-}$, the even and the odd part of
$\mathbf{u}$, given respectively by
\[
\mathbf{u}_{+}=\frac{1}{2}(\mathbf{u}+\hat{\mathbf{u}})\quad\text{and}%
\quad\mathbf{u}_{-}=\frac{1}{2}(\mathbf{u}-\hat{\mathbf{u}})
\]

The spaces $\bigwedge V$ and $\bigwedge V^{\ast}$ are identified with their
images in $\bigwedge H_{V}$ under the homomorphisms $i^{w}:\bigwedge V$
$\rightarrow\bigwedge H_{V}$ and $i_{w}:\bigwedge V^{\ast}\rightarrow\bigwedge
H_{V}$ by
\[
i^{w}(x_{1\ast}\wedge\ldots\wedge x_{r\ast})=(x_{1\ast}\oplus0)\wedge
\ldots\wedge(x_{r\ast}\oplus0)\equiv x_{1\ast}\wedge\ldots\wedge x_{r\ast}%
\]
and
\[
i_{w}(x_{1}^{\ast}\wedge\ldots\wedge x_{r}^{\ast})=(x_{1}^{\ast}\oplus
0)\wedge\ldots\wedge(x_{r}^{\ast}\oplus0)\equiv x_{1}^{\ast}\wedge\ldots\wedge
x_{r}^{\ast}.
\]
Then, for $u_{\ast},v_{\ast}\in\bigwedge V$ $\subset\bigwedge H_{V}$ and
$u^{\ast},v^{\ast}\in\bigwedge V^{\ast}\subset\bigwedge H_{V}$, we have
\begin{align}
&  \b<u_{\ast},v_{\ast}>=0,\nonumber\\
&  \b<u^{\ast},v^{\ast}>=0,\nonumber\\
&  \b<u^{\ast},v_{\ast}>=u^{\ast}(v_{\ast}). \label{EQ-14}%
\end{align}
Thus, $\bigwedge V$ and $\bigwedge V^{\ast}$ are totally isotropic subspaces
of $\bigwedge H_{V}$. But they are no longer maximal: $\b<\ ,\ >$ being
neutral, the dimension of a maximal totally isotropic subspace is $2^{2n-1}$,
whereas $\dim\bigwedge V$ $=\dim\bigwedge V^{\ast}=2^{n}$. For elements
$\mathbf{u}=u_{\ast}\wedge u^{\ast},\mathbf{v}=v_{\ast}\wedge v^{\ast}%
\in\bigwedge H_{V}$ with $u_{\ast}\in\bigwedge\nolimits^{r}V$ , $u^{\ast}%
\in\bigwedge\nolimits^{s}V^{\ast}$, $v_{\ast}\in\bigwedge\nolimits^{s}V$ , and
$v^{\ast}\in\bigwedge\nolimits^{r}V^{\ast}$ it holds
\[
\b<\mathbf{u},\mathbf{v}>=(-1)^{rs}u^{\ast}(v_{\ast})v^{\ast}(u_{\ast}).
\]

\noindent\textbf{Proposition.} There is the following natural isomorphism:
\[
\bigwedge H_{V}\simeq\bigwedge V\text{ }\widehat{\otimes}\bigwedge V^{\ast}
\]
where $\widehat{\otimes}$ denotes the graded tensor product. Moreover, being
$b$ a non-degenerate bilinear form on $V$ , it holds also
\[
\bigwedge H_{V}\simeq\bigwedge H_{bV}\text{ }\widehat{\otimes}\bigwedge
H_{-bV}.
\]

For a proof see, e.g., Greub \cite{Greub}. The first of the above isomorphisms
is given by the mapping $\bigwedge V$ $\widehat{\otimes}\bigwedge V^{\ast
}\rightarrow\bigwedge H_{V}$ by
\[
u_{\ast}\widehat{\otimes}u^{\ast}\mapsto i^{w}u_{\ast}\wedge i_{w}u^{\ast},
\]
for all $u_{\ast}\in\bigwedge V$ and $u^{\ast}\in\bigwedge V^{\ast}$. Under
this mapping, we can make the identification
\[
\bigwedge\nolimits^{r}H_{V}=\sum_{p+q=r}\bigwedge\nolimits^{p}V\text{
}\widehat{\otimes}\bigwedge\nolimits^{q}V^{\ast}\text{.}%
\]

\subsection{Contractions}

A left contraction $\mathbin\lrcorner:\bigwedge H_{V}\times\bigwedge
H_{V}\rightarrow\bigwedge H_{V}$ and a right contraction $\mathbin\llcorner
:\bigwedge H_{V}\times\bigwedge H_{V}\rightarrow\bigwedge H_{V}$ are
introduced in the algebra $\bigwedge H_{V}$ in the usual way (see, e.g.,
\cite{rodoliv2006}), i.e., by
\[
\b<\mathbf{u}\mathbin\lrcorner\mathbf{v},\mathbf{w}>=\b<\mathbf{v}%
,\tilde{\mathbf{u}}\wedge\mathbf{w}>,
\]
and
\[
\b<\mathbf{v}\mathbin\llcorner\mathbf{u},\mathbf{w}>=\b<\mathbf{v}%
,\mathbf{w}\wedge\tilde{\mathbf{u}}>,
\]
for all $\mathbf{u},\mathbf{v},\mathbf{w}\in\bigwedge H_{V}$. These operations
have the general properties
\begin{align*}
&  \mathbf{x}\mathbin\lrcorner\mathbf{y}=\mathbf{x}\mathbin\llcorner
\mathbf{y}=\b<\mathbf{x},\mathbf{y}>,\\
&  1\mathbin\lrcorner\mathbf{u}=\mathbf{u}\mathbin\llcorner=\mathbf{u}%
,\quad\mathbf{x}\mathbin\lrcorner=1\mathbin\llcorner\mathbf{x}=0,\\[1ex]
&  (\mathbf{u}\mathbin\lrcorner\mathbf{v})\hat{\ }=\hat{\mathbf{u}%
}\mathbin\lrcorner\hat{\mathbf{v}},\quad(\mathbf{u}\mathbin\llcorner
\mathbf{v})\hat{\ }=\hat{\mathbf{u}}\mathbin\llcorner\hat{\mathbf{v}},\\
&  (\mathbf{u}\mathbin\lrcorner\mathbf{v})\tilde{\ }=\tilde{\mathbf{u}%
}\mathbin\lrcorner\tilde{\mathbf{v}},\quad(\mathbf{u}\mathbin\llcorner
\mathbf{v})\tilde{\ }=\tilde{\mathbf{u}}\mathbin\llcorner\tilde{\mathbf{v}%
},\\[1ex]
&  \mathbf{u}\mathbin\lrcorner(\mathbf{v}\mathbin\lrcorner\mathbf{w}%
)=(\mathbf{u}\wedge\mathbf{v})\mathbin\lrcorner\mathbf{w,}\\
&  (\mathbf{u}\mathbin\llcorner\mathbf{v})\mathbin\llcorner\mathbf{w}%
=\mathbf{u}\mathbin\llcorner(\mathbf{v}\wedge\mathbf{w}),\\
&  (\mathbf{u}\mathbin\lrcorner\mathbf{v})\mathbin\llcorner\mathbf{w}%
=\mathbf{u}\mathbin\lrcorner(\mathbf{v}\mathbin\llcorner\mathbf{w}),\\[1ex]
&  \mathbf{x}\mathbin\lrcorner(\mathbf{u}\wedge\mathbf{v})=(\mathbf{x}%
\mathbin\lrcorner\mathbf{u})\wedge\mathbf{v}+\hat{\mathbf{u}}\wedge
(\mathbf{x}\mathbin\lrcorner\mathbf{v}),\\
&  (\mathbf{u}\wedge\mathbf{v})\mathbin\llcorner\mathbf{x}=\mathbf{u}%
\wedge(\mathbf{v}\mathbin\llcorner\mathbf{x})+(\mathbf{u}\mathbin\llcorner
\mathbf{x})\wedge\hat{\mathbf{v}},\\[1ex]
&  \mathbf{x}\wedge(\mathbf{u}\mathbin\lrcorner\mathbf{v})=\hat{\mathbf{u}%
}\mathbin\lrcorner(\mathbf{x}\wedge\mathbf{v})-(\hat{\mathbf{u}}%
\mathbin\llcorner\mathbf{x})\mathbin\lrcorner\mathbf{v}\\
&  (\mathbf{u}\mathbin\llcorner\mathbf{v})\wedge\mathbf{x}=(\mathbf{u}%
\wedge\mathbf{x})\mathbin\llcorner\hat{\mathbf{v}}-\mathbf{u}\mathbin\llcorner
(\mathbf{x}\mathbin\lrcorner\hat{\mathbf{v}})\\[1ex]
&  \mathbf{u}_{+}\mathbin\lrcorner\mathbf{v}=\mathbf{v}\mathbin\llcorner
\mathbf{u}_{+},\quad\mathbf{u}_{-}\mathbin\lrcorner\mathbf{v}=\hat{\mathbf{v}%
}\mathbin\llcorner\hat{\mathbf{u}}_{-}\\[1ex]
&  \mathbf{u}\wedge(\mathbf{v}\mathbin\lrcorner%
\hbox{\boldmath$\sigma$}%
)=(\mathbf{u}\mathbin\lrcorner\mathbf{v})\mathbin\lrcorner%
\hbox{\boldmath$\sigma$}%
,\\
&  (%
\hbox{\boldmath$\sigma$}%
\mathbin\llcorner\mathbf{u})\wedge\mathbf{v}=%
\hbox{\boldmath$\sigma$}%
\mathbin\llcorner(\mathbf{u}\mathbin\llcorner\mathbf{v}).
\end{align*}
Moreover, from Eqs.~\ref{EQ-14} we get (there are analogous results for right
contraction)
\begin{equation}
u_{\ast}\mathbin\lrcorner v_{\ast}=0\qquad\text{and}\qquad u^{\ast
}\mathbin\lrcorner v^{\ast}=0 \label{91}%
\end{equation}
for all $u_{\ast},v_{\ast}\in\bigwedge V$ and $u^{\ast},v^{\ast}\in\bigwedge
V^{\ast}$, so that for elements of the form $\mathbf{u}=u_{\ast}\wedge
u^{\ast}$ and $\mathbf{x}=x_{\ast}\oplus x^{\ast}$, it holds
\[
\mathbf{x}\mathbin\lrcorner\mathbf{u}=(x^{\ast}\mathbin\lrcorner u_{\ast
})\wedge u^{\ast}+\hat{u}_{\ast}\wedge(x_{\ast}\mathbin\lrcorner u^{\ast})
\]
For more details about the properties of left and right contractions, see,
e.g., \cite{Lounesto, rodoliv2006}

\subsection{Poincar\'{e} Automorphism (Hodge Dual)}

Define now the Poincar\'{e} automorphism\/(or Hodge dual\/) $\star:\bigwedge
H_{V}\rightarrow\bigwedge H_{V}$ by
\[
\star\mathbf{u}=\tilde{\mathbf{u}}\mathbin\lrcorner%
\hbox{\boldmath$\sigma$}%
,
\]
for all $\mathbf{u}\in\bigwedge H_{V}$. The inverse $\star^{-1}$ of this
operation is given by
\[
\star^{-1}\mathbf{u}=\tilde{%
\hbox{\boldmath$\sigma$}%
}\mathbin\llcorner\mathbf{\tilde{u}.}%
\]
The following general properties of the Hodge duality holds true
\begin{align*}
&  \star%
\hbox{\boldmath$\sigma$}%
=(-1)^{n},\quad\star^{-1}%
\hbox{\boldmath$\sigma$}%
=1\\
&  \b<\star\mathbf{u},\star\mathbf{v}>=(-1)^{n}\b<\mathbf{u},\mathbf{v}>\\
&  \star(\mathbf{u}\wedge\mathbf{v})=\tilde{\mathbf{v}}\mathbin\lrcorner
\star\mathbf{u}\\
&  \star^{-1}(\mathbf{u}\wedge\mathbf{v})=(\star^{-1}\mathbf{v}%
)\mathbin\llcorner\tilde{\mathbf{u}}\\
&  \star(\mathbf{u}\mathbin\llcorner\mathbf{v})=\tilde{\mathbf{v}}\wedge
\star\mathbf{u}\\
&  \star^{-1}(\mathbf{u}\mathbin\lrcorner\mathbf{v})=(\star^{-1}%
\mathbf{v})\wedge\tilde{\mathbf{u}}%
\end{align*}

For $\mathbf{x}=x_{\ast}\oplus x^{\ast}\in H_{V}\subset\bigwedge H_{V}$ we
have, since $x_{\ast}\mathbin\lrcorner e_{\ast}=0$ and $x^{\ast}%
\mathbin\lrcorner\theta^{\ast}=0$ that
\[
\star\mathbf{x}=(x^{\ast}\mathbin\lrcorner e_{\ast})\wedge\theta^{\ast
}-e_{\ast}\wedge(\theta^{\ast}\mathbin\llcorner x_{\ast}),
\]
and it follows that, for $u_{\ast}\in\bigwedge V$ $\subset\bigwedge H_{V}$ and
$u^{\ast}\in\bigwedge V^{\ast}\subset\bigwedge H_{V}$,
\[
\star u^{\ast}=(\tilde{u}^{\ast}\mathbin\lrcorner e_{\ast})\wedge\theta^{\ast
}=D_{\#}u^{\ast}\wedge\theta^{\ast}%
\]
and
\[
\star u_{\ast}=e_{\ast}\wedge(\theta^{\ast}\mathbin\llcorner\bar{u}_{\ast
})=e_{\ast}\wedge D^{\#}u_{\ast}%
\]
where we introduced the Poincar\'{e} isomorphisms\/ $D_{\#}:\bigwedge V^{\ast
}\rightarrow\bigwedge V$ and $D^{\#}:\bigwedge V$ $\rightarrow\bigwedge
V^{\ast}$ by (see, e.g. \cite{Greub}),
\[
D_{\#}u^{\ast}=\tilde{u}^{\ast}\mathbin\lrcorner e_{\ast}\qquad\text{and}%
\qquad D^{\#}u_{\ast}=\theta^{\ast}\mathbin\llcorner\bar{u}_{\ast}%
\]
For an element of the form $\mathbf{u}=u_{\ast}\wedge u^{\ast}$ with $u_{\ast
}\in%
{\displaystyle\bigwedge}
V$ and $u^{\ast}\in%
{\displaystyle\bigwedge}
V$ $^{\ast}$,
\[
\star\mathbf{u}=D_{\#}u^{\ast}\wedge D^{\#}u_{\ast}%
\]

\subsection{Differential Algebra Structure on $\bigwedge H_{V}$}

Provide the spaces $%
{\displaystyle\bigwedge}
V$ and $%
{\displaystyle\bigwedge}
V^{\ast}$ with structure of differential algebras, picking up linear operators
$\mathfrak{d}_{\ast}:%
{\displaystyle\bigwedge}
V$ $\rightarrow%
{\displaystyle\bigwedge}
V$ and $\mathfrak{d}^{\ast}:%
{\displaystyle\bigwedge}
V^{\ast}\rightarrow%
{\displaystyle\bigwedge}
V^{\ast}$, both satisfying
\begin{align*}
&  \mathfrak{d}^{2}=0\\
&  \mathfrak{d}\ \widehat{\ }+\widehat{\ }\ \mathfrak{d}=0\\
&  \mathfrak{d}(u\wedge v)=(\mathfrak{d}u)\wedge v+\hat{u}\wedge
(\mathfrak{d}v)
\end{align*}
with $\mathfrak{d}\equiv\mathfrak{d}_{\ast}$ or $\mathfrak{d}\equiv
\mathfrak{d}^{\ast}$ and $u,v\in%
{\displaystyle\bigwedge}
V$ or $u,v\in%
{\displaystyle\bigwedge}
V^{\ast}$, accordingly.

The pair $(\mathfrak{d}_{\ast},\mathfrak{d}^{\ast})$ induces a differential
operator $\mathfrak{d}$ on $%
{\displaystyle\bigwedge}
H_{V}$, defined as the unique\/ differential operator (with respect to the
grade involution of $%
{\displaystyle\bigwedge}
H_{V}$) satisfying
\[
\mathfrak{d}\ i^{w}=i_{\ast}^{w}\mathfrak{d}\qquad\text{and}\qquad
\mathfrak{d}\ i_{w}=i_{w}^{\ast}\mathfrak{d,}%
\]
where $i^{w}$ and $i_{w}$ are the canonical homomorphisms from $%
{\displaystyle\bigwedge}
V$ and $%
{\displaystyle\bigwedge}
V^{\ast}$ to $%
{\displaystyle\bigwedge}
H_{V}$. That is to say, for an element of $%
{\displaystyle\bigwedge}
H_{V}$ of the form $u=u_{\ast}\wedge u^{\ast}$ with $u_{\ast}\in%
{\displaystyle\bigwedge}
V$ and $u^{\ast}\in%
{\displaystyle\bigwedge}
V^{\ast}$, $\mathfrak{d}$ acts as
\[
\mathfrak{d}\mathbf{u}=(\mathfrak{d}_{\ast}u_{\ast})\wedge u^{\ast}+\hat
{u}_{\ast}\wedge(\mathfrak{d}^{\ast}u^{\ast})
\]

The simpler example of this construction is given by the left [or analogously
the right] contraction. Indeed,
\[
\mathfrak{d}_{\ast}=x^{\ast}\mathbin\lrcorner\quad\text{and}\quad
\mathfrak{d}^{\ast}=x_{\ast}\mathbin\lrcorner
\]
for some $x^{\ast}\in V$ and $x_{\ast}\in V^{\ast}$, are easily seen to be
differential operators on $%
{\displaystyle\bigwedge}
V$ and $%
{\displaystyle\bigwedge}
V^{\ast}$, respectively. Of course, they induce on $%
{\displaystyle\bigwedge}
H_{V}$ the operator
\[
\mathfrak{d}=\mathbf{x}\mathbin\lrcorner,
\]
with $\mathbf{x}=x_{\ast}\oplus x^{\ast}$.


\section{Clifford Algebra of a Hyperbolic Space: The Mother Algebra}

Introduce in $%
{\displaystyle\bigwedge}
H_{V}$ the Clifford product of a vector $\mathbf{x}\in H_{V}$ by an element
$\mathbf{u}\in%
{\displaystyle\bigwedge}
H_{V}$ by
\[
\mathbf{xu}=\mathbf{x}\mathbin\lrcorner\mathbf{u}+\mathbf{x}\wedge\mathbf{u}
\]
and extend this product by linearity and associativity to all of the space $%
{\displaystyle\bigwedge}
H_{V}$. The resulting algebra is isomorphic to the Clifford algebra
$\mathcal{C\ell}(H_{V})$ of the hyperbolic structure $H_{V}$ and will thereby
be identified with it.

We call $\mathcal{C\ell}(H_{V})$ the mother algebra\/ (or the hyperbolic
Clifford algebra\/) of the vector space $V$. The even and odd subspaces of
$\mathcal{C\ell}(H_{V})$ will be denoted respectively by $\mathcal{C\ell
}^{(0)}(H_{V})$ and $\mathcal{C\ell}^{(1)}(H_{V})$, so that
\[
\mathcal{C\ell}(H_{V})=\mathcal{C\ell}^{(0)}(H_{V})\oplus\mathcal{C\ell}%
^{(1)}(H_{V})
\]
and the same notation of the exterior algebra is used for grade involution,
reversion, and conjugation in $\mathcal{C\ell}(H_{V})$, which obviously
satisfy
\[
(\mathbf{uv})\hat{\ }=\hat{\mathbf{u}}\hat{\mathbf{v}},\qquad(\mathbf{uv}%
)\tilde{\ }=\tilde{\mathbf{v}}\tilde{\mathbf{u}},\qquad(\mathbf{uv})\bar
{\ }=\bar{\mathbf{v}}\bar{\mathbf{u}}%
\]

For vectors $\mathbf{x},\mathbf{y}\in H_{V}$, we have the relation
\[
\mathbf{xy}+\mathbf{yx}=2\b<\mathbf{x},\mathbf{y}>
\]
so that for the basis elements $\{%
\hbox{\boldmath$\sigma$}%
_{k}\}$ it holds
\begin{align*}
&
\hbox{\boldmath$\sigma$}%
_{k}%
\hbox{\boldmath$\sigma$}%
_{l}+%
\hbox{\boldmath$\sigma$}%
_{l}%
\hbox{\boldmath$\sigma$}%
_{k}=2\delta_{kl},\\
&
\hbox{\boldmath$\sigma$}%
_{n+k}%
\hbox{\boldmath$\sigma$}%
_{n+l}+%
\hbox{\boldmath$\sigma$}%
_{n+l}%
\hbox{\boldmath$\sigma$}%
_{n+k}=-2\delta_{kl},\\
&
\hbox{\boldmath$\sigma$}%
_{k}%
\hbox{\boldmath$\sigma$}%
_{n+l}=-%
\hbox{\boldmath$\sigma$}%
_{n+l}%
\hbox{\boldmath$\sigma$}%
_{k}.
\end{align*}
In turn, for the Witt basis $\{e_{k},\theta^{k}\}$, we have instead
\begin{align*}
&  e_{k}e_{l}+e_{l}e_{k}=0,\\
&  \theta^{k}\theta^{l}+\theta^{l}\theta^{k}=0,\\
&  \theta^{k}e_{l}+e_{l}\theta^{k}=2\delta_{l}^{k}.
\end{align*}

The Clifford product have the following general properties \begin{tabbing}
\hspace*{2cm} \= \\\kill
\>  $\mathbf{u}\leftcontract%
\hbox{\boldmath$\sigma$}%
=\mathbf{u}%
\hbox{\boldmath$\sigma$}%
,\quad%
\hbox{\boldmath$\sigma$}%
\rightcontract\mathbf{u}=%
\hbox{\boldmath$\sigma$}%
\mathbf{u}$\\[1ex]
\>  $\b<\mathbf{u},\mathbf{vw}>=\b<\tilde{\mathbf{v}}\mathbf{u},\mathbf{w}>
= \b<\mathbf{u}\tilde{\mathbf{w}},\mathbf{v}>$\\[1ex]
\>  $\mathbf{x}\wedge\mathbf{u}=\frac{1}{2}(\mathbf{xu}+\hat{\mathbf{u}%
}\mathbf{x}),\quad\mathbf{u}\wedge\mathbf{x}=\frac{1}{2}(\mathbf{ux}%
+\mathbf{x\hat{u}})$\\[1ex]
\>  $\mathbf{x}\leftcontract\mathbf{u}=\frac{1}{2}(\mathbf{xu}-\hat{\mathbf{u}%
}\mathbf{x}),\quad\mathbf{u}\rightcontract\mathbf{x}=\frac{1}{2}(\bu\bx-\bx\hat
{\bu})$\\[1ex]
\>  $\mathbf{x}\leftcontract(\mathbf{uv})=(\mathbf{x}\leftcontract\mathbf{u}%
)\mathbf{v}+\hat{\mathbf{u}}(\mathbf{x}\leftcontract\mathbf{v})$ \\[1ex]
\>  $(\mathbf{uv})\rightcontract\mathbf{x}=\mathbf{u}(\mathbf{v}\rightcontract
\mathbf{x}+(\mathbf{u}\rightcontract\mathbf{x})\hat{\mathbf{v}}$\\[1ex]
\>  $\mathbf{x}\wedge(\mathbf{uv})=(\mathbf{x}\leftcontract\mathbf{u})\mathbf{v}%
+\hat{\mathbf{u}}(\mathbf{x}\wedge\mathbf{v})=(\mathbf{x}\wedge\mathbf{u}%
)\mathbf{v}-\hat{\mathbf{u}}(\mathbf{x}\leftcontract\mathbf{v})$\\[1ex]
\>  $(\mathbf{uv})\wedge\mathbf{x}=\mathbf{u}(\mathbf{v}\wedge\mathbf{x}%
)-(\mathbf{u}\rightcontract\mathbf{x})\hat{\mathbf{v}}=\mathbf{u}(\mathbf{v}%
\rightcontract\mathbf{x})+(\mathbf{u}\wedge\mathbf{x})\hat{\mathbf{v}}$\\[1ex]
\>  $\star\mathbf{u}=\tilde{\mathbf{u}}%
\hbox{\boldmath$\sigma$}%
,\quad\star^{-1}\mathbf{u}=\tilde{%
\hbox{\boldmath$\sigma$}%
}\tilde{\mathbf{u}}$\\[1ex]
\>  $\star(\mathbf{uv})=\tilde{\mathbf{v}}(\star\mathbf{u}),\quad\star
^{-1}(\mathbf{uv})=(\star^{-1}\mathbf{v})\mathbf{u}$%
\end{tabbing}Moreover, for an element of the form $\mathbf{u}=u_{\ast}\wedge
u^{\ast}$, with $u_{\ast}\in%
{\displaystyle\bigwedge}
V$ and $u^{\ast}\in%
{\displaystyle\bigwedge}
V^{\ast}$,
\[
\mathbf{xu}=(x^{\ast}u_{\ast})\wedge u^{\ast}+\hat{u}_{\ast}\wedge(x_{\ast
}u^{\ast})
\]
and it is also important to note that the square of the volume $2n$-vector $%
\hbox{\boldmath$\sigma$}%
$ satisfy
\[%
\hbox{\boldmath$\sigma$}%
^{2}=1
\]

\noindent\textbf{Proposition 2}. There is the following natural isomorphism
\[
\mathcal{C\ell}(H_{V})\simeq\mathop{\mathrm{End}}(%
{\displaystyle\bigwedge}
V).
\]
In addition, being $b$ a non-degenerate symmetric bilinear form on $V$, it
holds also
\[
\mathcal{C\ell}(H_{V})\simeq\mathcal{C\ell}(H_{bV})\simeq\mathcal{C\ell
}(V,b)\text{ }\widehat{\otimes}\text{ }\mathcal{C\ell}(V,-b).
\]

\noindent\textbf{Proof. }The first isomorphism is given by the extension to
$\mathcal{C\ell}(H_{V})$ of the Clifford map\/ $\varphi:H_{V}\rightarrow
\mathop{\mathrm{End}}(%
{\displaystyle\bigwedge}
V$ $)$ by $\mathbf{x}\mapsto\varphi_{x}$, with
\[
\varphi_{x}(u_{\ast})={\frac{1}{\sqrt{2}}}\left(  x^{\ast}\mathbin\lrcorner
u_{\ast}+x_{\ast}\wedge u_{\ast}\right)  ,
\]
for all $u_{\ast}\in%
{\displaystyle\bigwedge}
V$. The second isomorphism, in turn, is induced from the Clifford map
\[
x_{\ast}\oplus x^{\ast}\mapsto x_{+}\widehat{\otimes}1+1\widehat{\otimes}%
x_{-},
\]
with
\[
x_{\pm}={{\frac{1}{\sqrt{2}}}}(b^{\ast}x^{\ast}\pm x_{\ast}).
\]

\noindent\textbf{Corollary.} The even and odd subspaces of the hyperbolic
Clifford algebra are
\[
\mathcal{C\ell}^{(0)}(H_{V})\simeq\mathop{\mathrm{End}}(%
{\displaystyle\bigwedge\nolimits^{(0)}}
V)\oplus\mathop{\mathrm{End}}(%
{\displaystyle\bigwedge\nolimits^{(1)}}
V)
\]
and
\[
\mathcal{C\ell}^{(1)}(H_{V})\simeq L(%
{\displaystyle\bigwedge\nolimits^{(0)}}
V,%
{\displaystyle\bigwedge\nolimits^{(1)}}
V)\oplus L(%
{\displaystyle\bigwedge\nolimits^{(1)}}
{\displaystyle\bigwedge}
,%
{\displaystyle\bigwedge\nolimits^{(0)}}
V)
\]
where $L(V,W)$ denotes the space of the linear mappings from $V$ to $W$ and $%
{\displaystyle\bigwedge\nolimits^{(0)}}
V$ and $%
{\displaystyle\bigwedge\nolimits^{(1)}}
V$ denote respectively the spaces of the even and of \ odd elements of $%
{\displaystyle\bigwedge}
V$ .

\noindent\textbf{Remark.} Recalling the definition of the second order
hyperbolic structure, $H_{V}^{2}$, it follows from Proposition~2 that
\[
\mathcal{C\ell}(H_{V}^{2})\simeq\mathop{\mathrm{End}}(\mathcal{C\ell}(H_{V})
\]
The algebra $\mathcal{C\ell}(H_{V}^{2})$ may be called \textit{grandmother}
algebra\/ of the vector space $V$.\hfill\ 

\section{Ideals in $\mathcal{C\ell}(H_{V})$. Superfields}

It is a well known result that minimal ideals in Clifford algebras are the
spaces representing spinors (for more details see, e.g., \cite{walrod,
rodoliv2006}). Given a pair of dual basis $\{e_{1},\ldots,e_{n}\}$ in $V$ and
$\{\theta^{1},\ldots,\theta^{n}\}$ in $V^{\ast}$ the $n$-form volume element
$\theta^{\ast}=\theta^{1}\wedge\ldots\wedge\theta^{n}$. Taking into account
Eq.(\ref{91}) (and the analogous ones for the right contraction) we
immediately realize that $%
{\displaystyle\bigwedge}
V^{\ast}$ (the space of multiforms) is a minimal left ideal of the hyperbolic
Clifford algebra $\mathcal{C\ell}(H_{V})$, i.e., we have \cite{quinrod}%
\[%
{\displaystyle\bigwedge}
V^{\ast}=\mathcal{C\ell}(H_{V})\theta^{\ast}.
\]

A typical element of $\psi\in%
{\displaystyle\bigwedge}
V^{\ast}$ which is a \textit{spinor} in $\mathcal{C\ell}(H_{V})$ is then
written as:%
\[
\psi=s+v_{\mu}\theta^{\mu}+\frac{1}{2}f_{\mu\nu}\theta^{\mu}\theta^{\nu
}+\ldots+p\theta^{1}\wedge\ldots\wedge\theta^{n},
\]
where $s,v_{\mu},f_{\mu\nu},\ldots,p\in\mathbb{R}$.

The algebraic structure described above can be transferred for spin manifolds
through the construction of the Hyperbolic Clifford bundle $\mathcal{C\ell
}(H(M)):=\mathcal{C\ell}(TM\oplus T^{\ast}M,\b<\ ,\ >)$ and the elements
of\footnote{Of course, now $\theta^{\ast}\in\sec%
{\displaystyle\bigwedge}
T^{\ast}M\hookrightarrow\sec\mathcal{C\ell}(H(M))$ refers to a volume element
in $M$ defined by a section of the coframe bundle, i.e., $\{\theta^{i}%
\}\in\sec F(M)$.} $\mathcal{C\ell}(H(M))\theta^{\ast}$ are Witten
superfields\cite{witten}. It is opportune to recall here that in
\cite{walrod,moro} we showed that Dirac spinors can be rigorously represented
as equivalence classes of (even) sections of the Clifford bundle of
differential forms $\mathcal{C\ell}(T^{\ast}M,\mathtt{g})$ of a Lorentzian
spacetime, and in this sense the electron field is a kind of superfield. More
details on this issue can be found in \cite{rodoliv2006}.

\section{Conclusions}

In this paper we introduced and studied with details the Clifford algebra
$\mathcal{C\ell(}H_{V})$ of multivecfors of a hyperbolic space $H_{V}$, and
give the geometrical meaning of those objects. We recall that $\mathcal{C\ell
(}H_{V})$ is the algebra of endomorphisms of $%
{\displaystyle\bigwedge}
V$, something that make mutivecfors really important objects, although they
are not yet general enough to represent the operators we use in differential
geometry.\footnote{The general operator need in the study of differential
geometry are the extensors (and extensor fields) of the hyperbolic space that
will be presented in a series of forthcoming papers.} Besides that we showed
that $%
{\displaystyle\bigwedge}
V^{\ast}$ is a particular minimum ideal in $\mathcal{C\ell(}H_{V})$ whose
elements are representatives of spinors and which describe the
\textit{algebraic properties} of Witten superfields. We end with the
observation that Clifford and Grassmann algebras can be generalized in several
different ways and in particular the approach in \cite{rovaz} that deals with
Peano spaces seems particularly interesting to be further studied concerning
construction of hyperbolic structures. We will return to this issue in a
forthcoming paper.

\end{document}